# On the Physical Meaning of the Geometric Factor and the Effective Thickness in the Montgomery Method

F. S. Oliveira[1], L. M. S. Alves[2], M. S. da Luz[3], E. C. Romão[4], and C. A. M. dos Santos[4]

[1]*Instituto de Física "Gleb Wataghin", Universidade Estadual de Campinas, Campinas - SP, 13083-970, Brazil*

[2]*Instituto Federal de Educação, Ciência e Tecnologia Catarinense, Araquari - SC, 89245-000, Brazil*

[3]*Instituto de Ciências Tecnológicas e Exatas, Universidade Federal do Triângulo Mineiro, Uberaba - MG, 38025-180, Brazil*

[4]*Escola de Engenharia de Lorena, Universidade de São Paulo, Lorena - SP, 12602-810, Brazil*

## Abstract

Simulations carried out with COMSOL software in order to study the electrical resistivity of rectangular samples are reported. The comparison of the results with the four-probe method allows to understand the meaning of the geometric factor ($H$) and the effectiveness thickness ($E$) defined in the Montgomery method.

1) **Introduction**

The Montgomery and the van der Pauw methods have been extensively used to characterize the electrical resistivity ($\rho$) of isotropic and anisotropic conductors [1,2]. These methods have many technical advantages with regard to the preparation of probes in samples with non-uniform geometric and/or millimetric size [3], which avoid contact shot-circuit in comparison to the conventional four-probe method [4].

The Montgomery method involves attaching contacts to the opposite sides of a rectangular flat sample for measuring the electrical resistance between these contacts. The sample is then rotated, and the resistance is measured again [2]. By using the measured resistances, the electrical resistivity of the sample can be determined [2]. Furthermore, the van der Pauw method is similar but is used for samples with more complex geometries [1]. Thus, the key advantages of both methods are the ability to accurately measure the electrical properties of samples with complex geometries, such as irregularly shaped samples and anisotropic conductors [1].

In addition, a simple equation to calculate electrical resistivity has been published long ago by Montgomery [2] as

$$\rho = R_1 H_1 E, \qquad (1)$$

where $E$ is the effective thickness and the geometric parameter

$$H_1 \approx \frac{\pi}{8} \sinh\left(\pi \frac{L_2}{L_1}\right), \qquad (2)$$

provides the intrinsic electrical resistivity of an isotropic sample as a function of the electrical resistance ($R_1$), which is measured using four electrical probes placed in the corners of rectangular blocks, with $L_1$ and $L_2$ sides.

An external electrical current $I$ flows between two consecutive corners while the drop of the electric potential $V_1$ is measured between the two remanent corners on the opposite side (for instance, see Fig. 1 and 2 of the reference [3]). The ratio $V_1/I_1$ provides the sample electrical resistance ($R_1$). The geometrical parameters $H_1$ and $E$ in Equation (1) are related to the surface aspect ratio of a rectangular block ($L_2/L_1$) and its thickness, respectively [2]. Although the Montgomery has been published more than 50 years ago, there is a lack of physical meaning for these two geometrical parameters.

Unfortunately, the experimental investigation of the $H_1$ and $E$ in the Montgomery method is limited because requires the preparation and the measurement of a large number of samples with several sizes and different conducting materials, which is not a simple experimental task. On the other hand, the numerical simulations using the finite element method have some advantages because they can reproduce easily the experiments in rectangular blocks with many different sizes, without spending too much time. Furthermore, this numerical approach allows a systematic and deep analysis of the equipotential and electric field profiles in a conducting material under an external electrical current flow, which is not generally obtained in the transport measurements. The powerful of the finite element analysis performed using the COMSOL software can be observed in our recent publication [4].

Thus, in this work, we carried out a numerical modeling study to broaden the comprehension of the physical meaning of the geometric factor and effective thickness in the Montgomery method [2, 3]. We have noticed that the comparison of the Montgomery method with the conventional four-probe method provides insights into the definition of the $H_1$ term. Furthermore, the analysis of the electrical resistivity of 3D samples clarifies the meaning of the effective thickness ($E$) of rectangular conducting blocks.

## 2) Methodology

2D isotropic conducting simulated samples with $L_1$ and $L_2$ dimensions are studied, using COMSOL Multiphysics software [5]. The length $L_1$ was fixed at 1 m and $L_2$ was varied within the range $L_2/L_1$ from 0.01 to 8.

COMSOL has been already used to describe the behavior of electrical resistivity of several samples with rectangular shape typically used in the Montgomery method [4] and is applied in many areas, such as electric field, plasma physics, etc [6-9].

The squared COMSOL mesh grid was used to perform the finite element calculations in the 2D modeling using finite elements in quadrilateral shape. The lowest size for the MFE (Maximum Finite Element) of the mesh grid was defined as 0.01 m, which means that the minimum $L_2/L_1 = 0.01$ aspect ratio corresponds to the 1D boundary in this study.

In Fig. 1(a) the external applied current ($I_1$) of 1 A was chosen as the boundary condition using L-shape corners of 0.01 m × 0.01 m at point $A$ such as the electrical current terminal, while the same type of contact in the corner at point $B$ was used as the ground terminal. The voltage drop ($V_1$) induced between the corners placed at the points $C$ and $D$, due to a steady electrical current flow through points $A$ and $B$, was simulated using isotropic samples with electrical resistivity of 1 Ω m. Color pallet in the Fig. 1(a) highlights the boundaries in which the equipotential lines related to points $C$ and $D$ points are found in the sample.

The third dimension, the thickness ($L_3$), was studied in order to investigate the effective thickness ($E$) in the Montgomery method [2, 3]. Finite element modeling simulations were performed in 3D systems of different dimensions using standard triangular mesh grids in extra fine size. In Fig. 1(*b*) is shown the dropped voltage simulated between *C* and *D* corners induced by 1 A external electrical current applied between the A and B corners. The results were analyzed for samples with $L_2 \geq L_1$ and thickness ranging between 0.01 m and 8 m.

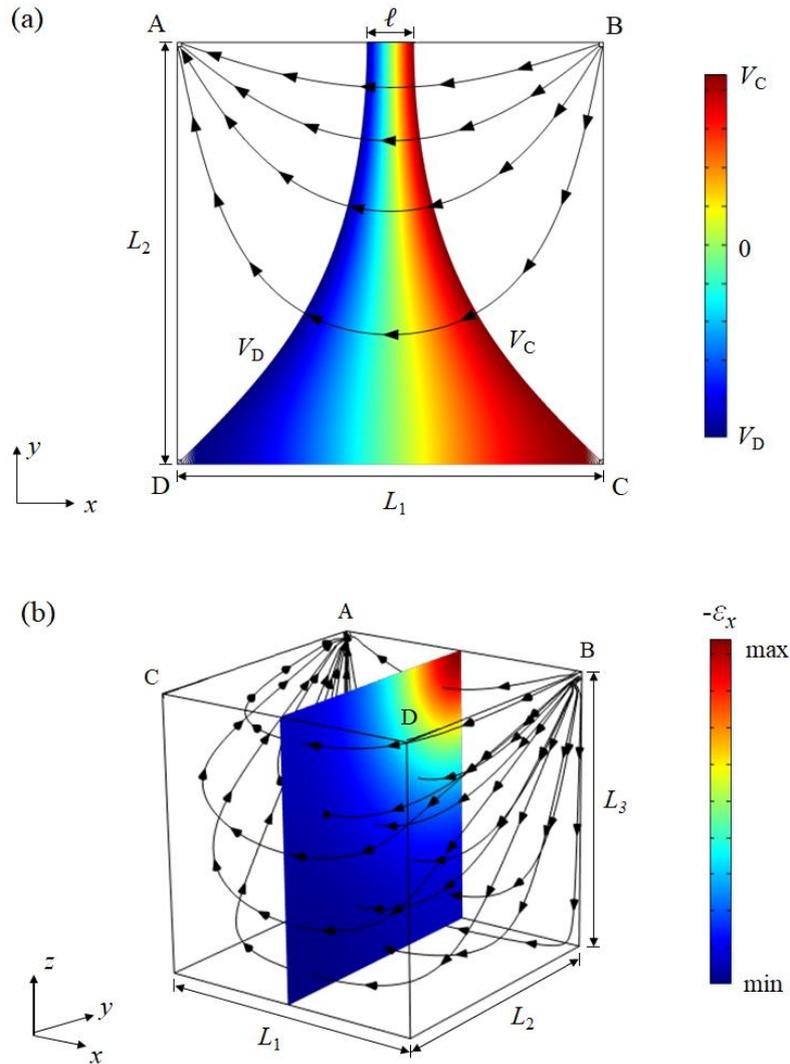

**Figure 1** – (*a*) Typical 2D simulation for a rectangular sample with $L_1$ and $L_2$ sides. Voltage is applied at the corners A and B in the way they provide the voltage profile shown by different colors between C and D points, whose values depend on the electrical resistivity of the sample. $\ell$ represents the distance between the points along the AB line where the voltage drop is equal to $V_1 = V_D - V_C$. In (*b*) a typical 3D simulation is shown, in which $L_3$ is the thickness of the sample. The color pallet represents the electric field in the middle section of the sample. In both figures, the arrows represent the electric field ($\varepsilon_x$) lines.

### 3) Results and discussion

#### 3.1) Interpretation for the geometric factor ($H_1$)

Fig. 2 displays the schematic samples used in the conventional four-probe and the Montgomery methods.

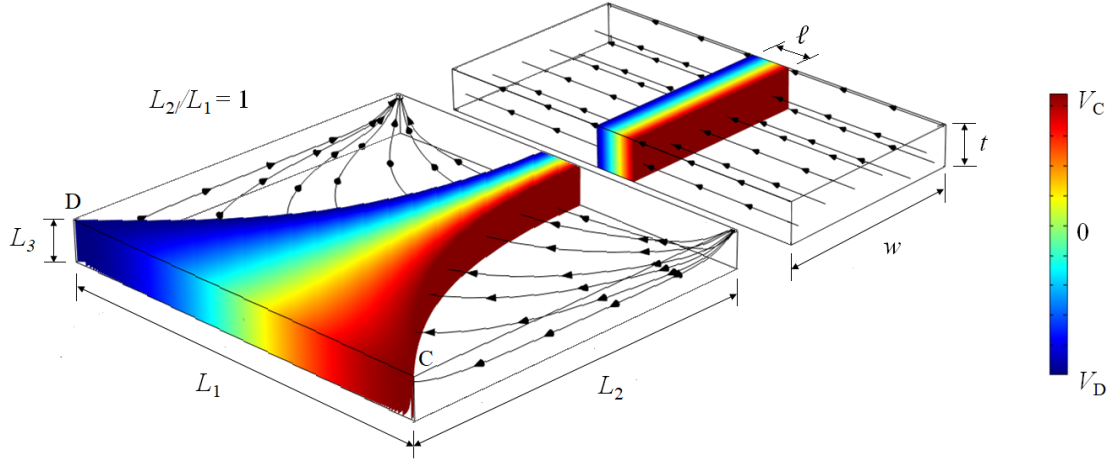

**Figure 2** – Equivalence between samples used in the conventional four-probe (right) and in the Montgomery methods (left). The dimensions are described in the text.

The conventional four-probe method uses a sample with four aligned contacts, which provides the electrical resistivity given by

$$\rho = \frac{V_1}{I_1} \frac{wt}{\ell}, \qquad (3)$$

where $w$ is the sample width, $\ell$ is the distance between the voltage contacts, and $t$ is the sample thickness.

Making a rectangular sample with $E = L_3 = t$, as drawn in Fig. 2, taking $H_1$ given by Equation (2), and comparing the electrical resistivity in the Montgomery method (Equation 1) with the four-probe method (Equation 3), it is easy to show that

$$\frac{w}{\ell} = H_1 \approx \frac{\pi}{8} \sinh\left(\pi \frac{L_2}{L_1}\right). \qquad (4)$$

This is interesting since it provides a direct physical meaning for the geometrical parameter $H_1$ of the Montgomery method, which is related to the ratio between the sample width and the distance between the voltage contacts in the four-probe method (a similar meaning can be obtained for the geometrical parameter $H_2$ [2,3]).

Simulations similar to the Fig. 1(*a*) were performed in several isotropic rectangular conducting sheets with different aspect ratios $L_2/L_1$ in order to deeper understand the meaning of the geometrical parameter $H_1$ described above. Some results are displayed in Fig. 3 for different sample dimensions.

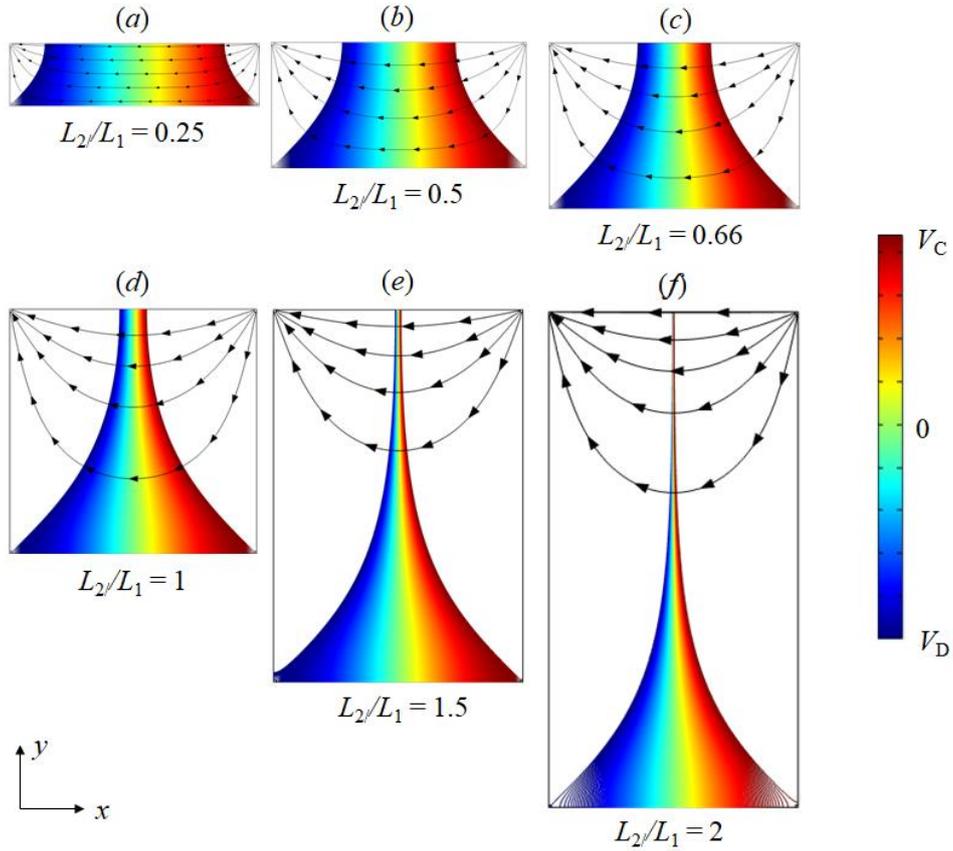

**Figure 3** – (*a*) - (*f*) 2D simulations for some different aspect ratios. As $L_2/L_1$ increases, the equivalent distance between the voltage contacts $\ell$, defined in Fig. 2, strongly decreases.

Taking the simulation data for the electric voltage along A-B and C-D segments, it is possible to find the distance $\ell$ very precisely. Fig. 4(*a*) displays these results and the way to determine $\ell$ for a 2D sample with $L_1 = L_2 = 1$ m. In Fig. 4(*b*) is shown the results of $V_1 (= V_{CD})$, $\ell$, and $w$ for samples with different aspect ratios $L_2/L_1$.

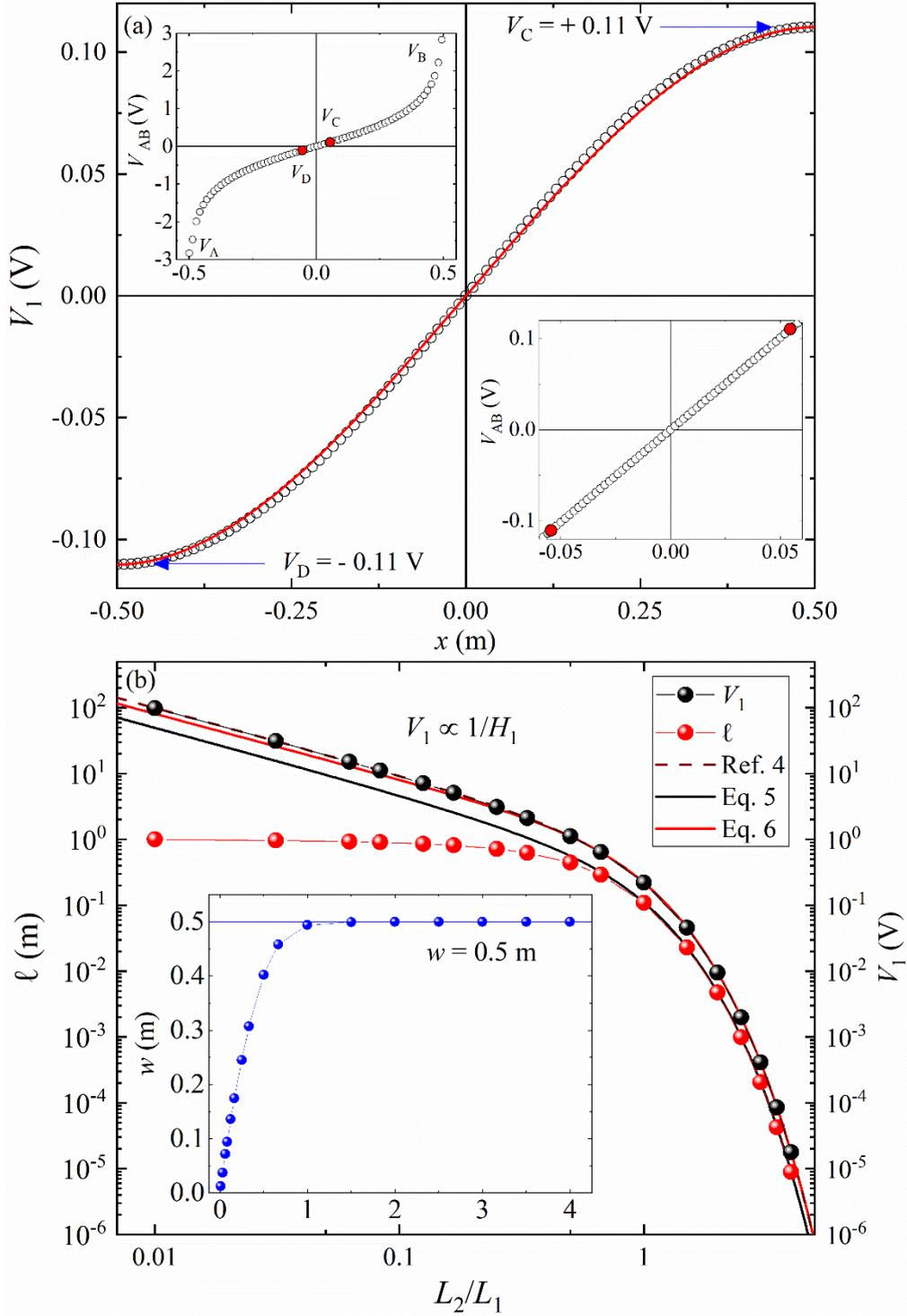

**Figure 4** – (*a*) Electric potential along A-B (upper inset) and C-D (main panel) segments for a 2D sample with $L_1 = L_2 = 1$ m. The red full line is the fit expected for the $V_1$ behavior, which predicts the saddle points at corners C and D [4]. The $V_C$ and $V_D$ values at these points are indicated by the horizontal blue arrows. The lower inset shows how the $\ell$ value is determined, which corresponds to the segment in the A-B direction where the voltage is equal to $V_1 = V_{CD} = V_D - V_C$. In (*b*) is shown the behavior of $V_1$ and $\ell$ as a function of the ratio $L_2/L_1$. Fitting lines are given by Equations (5), (6), and (4) in reference 4. Insert displays the behavior of $w$, which is equal $L_1/2$ above the limit $L_2/L_1 > 1$.

Results show that the $V_1$ follows the expected result given by Equation (6) of reference 5. On the other hand, surprising are the behaviors of $\ell$ and $w$. $\ell$ has similar dependence as $H_1$, given by

$$\ell = L_1 \frac{4}{\pi} \text{csch}\left(\pi \frac{L_2}{L_1}\right), \qquad (5)$$

at least in the $L_2/L_1 > 1$ limit (see black line and red circle symbols in Fig. 4 (b)).

In addition, by comparing this result with Equation (4), one can notice that $w = L_1/2$ is an implication for all samples in this limit (see inset of Fig. 6(b)), no matter the dimensions of the samples.

### 3.2) Interpretation of the effective thickness ($E$)

Fig. 5 displays simulations of 3D samples showing the electric field lines (see arrows) and the profiles at the center of the rectangular blocks with different thickness $L_3$.

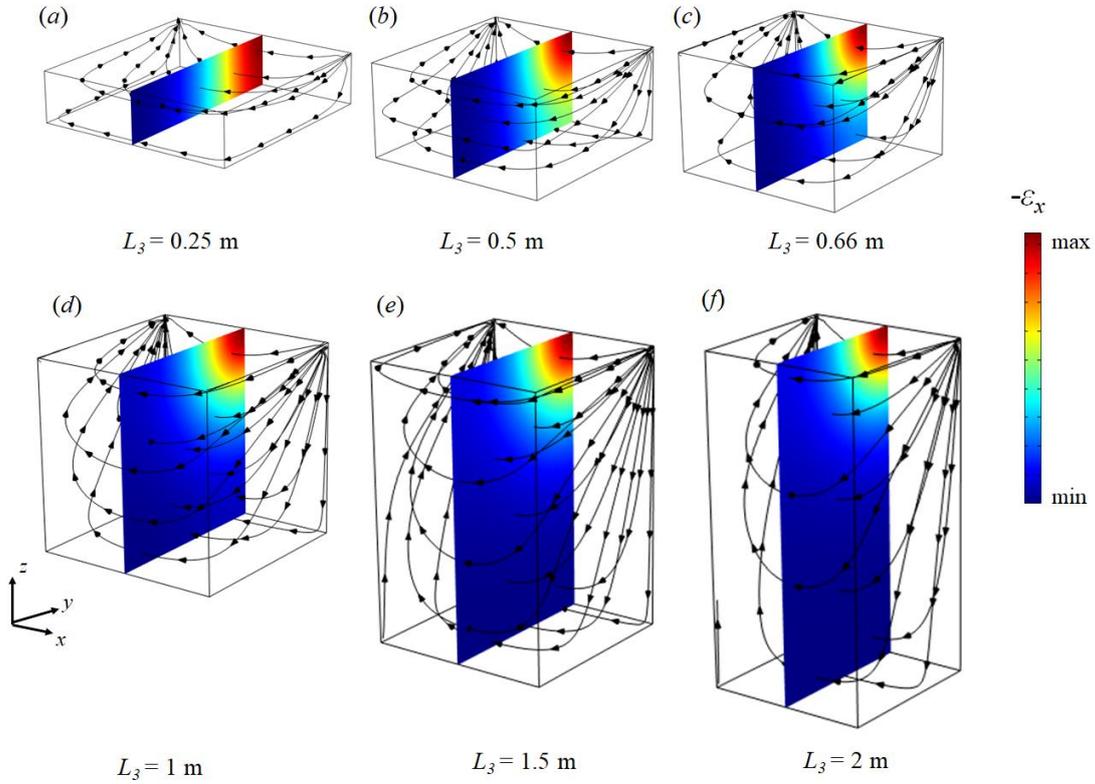

**Figure 5** – (*a*) - (*f*) Behavior of the electric field in 3D rectangular samples with different thickness $L_3$. Arrows show the direction of the electric field. The colored profiles display the electric field vector at the center for each sample, which is directed along the x-axis ($\varepsilon_x$).

It is observed that the profile of thin samples is different than the thicker ones. Thin samples have homogeneous electric field profile along the thickness (colored lines are almost vertical), although, thicker blocks have inhomogeneous profile, which tend to saturate the value of the electric fields in the bottom of the block (dark blue color region).

Taking the simulation data, it is possible to study the behavior of the electric voltage and the electric field for several samples. Fig. 6 displays the behavior of $V_1$ as a function of the thickness $L_3$ for samples with different $L_2/L_1$ ratios.

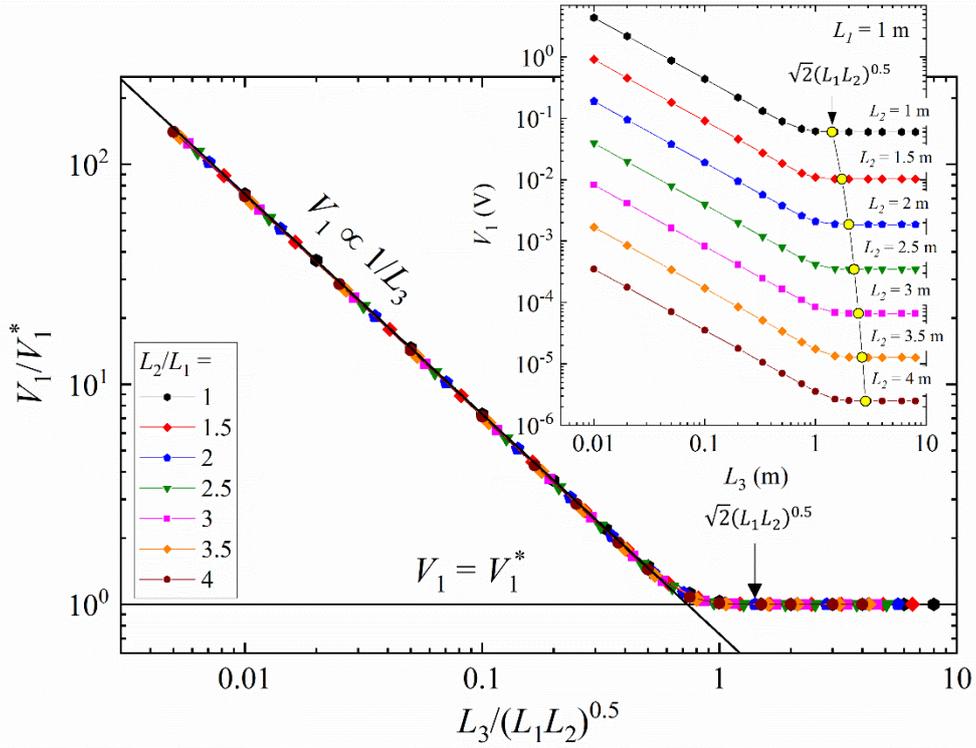

**Figure 6** – $V_1$ as a function of the thickness $L_3$ for samples with different $L_2/L_1$ ratios. Two behaviors are clearly noticed, one linear and another constant, which are related to the Equations (6), for $E = L_3$, and (8), respectively. The constant behavior of $V_1$ is reached at $L_3 = (2L_1L_2)^{1/2}$, which are represented by the yellow solid circles shown in the insert. The main panel displays the universal scaling taking $V_1/V_1^*$ as a function of $L_3/(L_1L_2)^{1/2}$, where $V_1^*$ is the saturation of $V_1$ for each sample.

In fact, the linear behaviors and the saturations observed in Fig. 6 are related to the Equation (6) reported previously [4], which is given by

$$V_1 = \frac{8}{\pi}\rho I \frac{1}{\sinh(\pi L_2/L_1)}\frac{1}{E}, \qquad (6)$$

in the $L_2/L_1 \geq 1$ limit.

The linear behavior of Fig. 6 is found taking $E = L_3$ and the saturation of $V_1$ occurs at the maximum effective thickness (yellow full circles), which is obtained by

$$E^* = \frac{\sqrt{2}}{2} (L_1 L_2)^{1/2}, \qquad (7)$$

in agreement with the previous works [2].

These two behaviors physically represent the thin sample limit and sample with maximum effective thickness, above which the electrical resistance of the sample does not change with increasing its thickness.

Additionally, such as the behaviors of $V_1$ in the insert of Fig. 6 are similar for all samples, it is possible to make a universal scaling taking $V_1/V_1^*$ as a function of $L_1$, $L_2$, and $L_3$, where $V_1^*$ is the saturation of $V_1$ for each sample, which is given by

$$V_1^* = \frac{8}{\pi} \rho I \frac{1}{\sinh(\pi L_2/L_1)} \frac{\sqrt{2}}{(L_1 L_2)^{1/2}}. \qquad (8)$$

Dividing Equation (6) by Equation (7) and taking the limit $E = L_3$, it is easy to notice that $V_1/V_1^*$ must scale as a function of $L_3/(L_1 L_2)^{1/2}$. The main panel of Fig. 6 displays this scaling, which shows an excellent collapse for all data.

Finally, further analysis of this collapse shows that the function which describes this universal behavior seems to be simple and is related to the normalized effective thickness reported previously [2-4]. This study is underway and the results will be reported elsewhere [10].

## 4) Conclusions

Numerical simulations using COMSOL software allowed us to understand the geometrical parameters ($V_1$ and $E$) of the Montgomery method [2]. The simulations show unambiguously that the geometrical parameter $H_1$ is related to the width and the distance between voltage probes of an equivalent sample measured in the conventional four-probe method. Furthermore, the simulations tuning the thickness of the rectangular samples provided important insights into the effective thickness in the Montgomery method. Two behaviors are clearly observed in the results, one related to the thin samples limit, $E = L_3$, and another due to a saturation of the voltage $V_1$ at $L_3 = \sqrt{2}/2 \, (L_1 L_2)^{1/2}$, above which the sample resistance is independent of the thickness. The results are in excellent agreement with the previous results [2-4].

**Acknowledgments**

F. S. Oliveira is a post-doc at UNICAMP (FAPESP Grant No. 2021/03298-7). M. S. da Luz is a CNPq fellow (Proc. 311394/2021-3).

**Conflict of Interest**

The authors have no conflicts to disclose.

**References**

[1] L. J. van der Pauw, "A method of measuring specific resistivity and Hall effect of discs of arbitrary shape", Philips Res. Rep. 13, 1 (1958).

[2] H. C. Montgomery, "Method for measuring electrical resistivity of anisotropic materials," J. Appl. Phys. 42, 2971 (1971).

[3] F. S. Oliveira, R. B. Cipriano, F. T. da Silva, E. C. Romão, and C. A. M. dos Santos, "Simple analytical method for determining electrical resistivity and sheet resistance using the van der Pauw procedure," Sci. Rep. 10, 16379 (2020) and references therein.

[4] L. M. S. Alves, F. S. Oliveira, E. C. Romão, M. S. da Luz and C. A. M. dos Santos, "A simple and precise way to determine electrical resistivity of isotropic conductors: Simplifying the four-probe method", AIP Advances 13, 035134 (2023).

[5] COMSOL™ 5.1 Multiphysics, Reference Manual, 2015.

[6] J. Lin, P. Y. Wong, P. Yang, Y. Y. Lau, W. Tang, and P. Zhang, "Electric field distribution and current emission in a miniaturized geometrical diode," J. Appl. Phys. 121, 244301 (2017).

[7] A. I. Sharshir, S. A. Fayek, Amal. F. Abd El-Gawad, M. A. Farahat, M.I. Ismail, and Mohamed Mohamady Ghobashy, "Experimental investigation of E-beam effect on the electric field distribution in cross-linked polyethy-lene/ZnO nanocomposites for medium-

voltage cables simulated by COMSOL Multiphysics," J. Anal. Sci. Technol. 13, 16 (2022).

[8] P. Song, Q. Song, Z. Yang, G. Zeng, H. Xu, X. Li, and W. Xiong, "Numerical simulation and exploration of electrocoagulation process for arsenic and antimony removal: Electric field, flow field, and mass transfer studies," J. Environ. Manage. 228, 336 (2018).

[9] E. Turkoz, M. Celik, "AETHER: A simulation platform for inductively coupled plasma", Journal of Computational Physics 286 (2015) 87–102.

[10] C. A. M. dos Santos *et al.* In preparation (2023).